# Analytical analysis of the decrease in the focal intensity of a femtosecond laser pulse caused by an imperfect surface of compressor diffraction gratings

E.A. Khazanov

*Federal Research Center A.V. Gaponov-Grekhov Institute of Applied Physics of the Russian Academy of Sciences (IAP RAS)*

*An analytical expression for the focal intensity of a laser pulse was obtained for an asymmetric out-of-plane compressor with gratings of arbitrary surface shape. The focal intensity is most strongly affected by the linear angular chirp caused by the spatial shift of different frequencies on the second and third gratings. The chirp can be eliminated by simply rotating the fourth grating by an optimal angle, which significantly reduces the requirements for the grating quality. It is shown that the decrease in the focal intensity depends on the product of grating surface rms and pulse spectrum bandwidth. With low-quality gratings, spectrum narrowing would not reduce focal intensity; contrariwise, it may even slightly increase it.*

## 1. Introduction

Nearly all high-power femtosecond lasers are based on the CPA [1] or OPCPA [2] architecture. The compressor is a key element of such lasers, especially multi-PW ones. The vast majority of lasers use a classical Treacy compressor (TC) [3], consisting of two identical pairs of diffraction grating. The gratings in each pair are parallel and the pairs mirror each other, i.e. it is a fully symmetric compressor (SC). The beam in the compressor propagates only in the horizontal plane (in the diffraction plane). In other words, the wave vector at any compressor point and at any frequency lies in the xz plane, so the TC is a plane compressor (PC). The TC features have been studied in many works, e.g., [4-7]. The TC is characterized only by three parameters: the distance between the gratings $L$, the groove density $N$, and the angle of incidence $\alpha$. In recent years, two directions of TC modification with more parameters have been discussed in the literature, see Table 1. The first direction – an asymmetric compressor (AC) – is based on the rejection of symmetry. The second one – an out-of-plane compressor (OC) – is based on the rejection of plane geometry.

An AC, in which the distances $L$ between the gratings in the pairs are different, was proposed and numerically studied in [8, 9]. An important AC property is the smoothing of fluence fluctuations at the output, which can significantly reduce the probability of optical breakdown of the fourth grating, adaptive mirror and focusing parabola. An analytical theory of AC, in which the grating pairs may differ not only in $L$, but also in $N$ and $\alpha$ (Fig. 1a), was constructed in [10]. It was shown theoretically that no compressor asymmetry reduces the focal intensity, and this conclusion is also valid for a compressor with one pair of gratings, which is a particular case of AC [11-13] (Table 1). Note also that the compressor asymmetry by no means excludes the possibility of pulse post-compression [14, 15], as shown in [16], it even expands the potential for its applications thanks to the above-mentioned beam smoothing [17]. The pulse self-compression after AC is discussed in [18].

The angle of incidence $\gamma$ in the plane normal to the diffraction plane is different from zero (Fig. 1b) in the OC [19-26]. The OC is used, for example, for spectral beam combining [24] and for compressing narrowband pulses [25]. Both multilayer dielectric and gold gratings in the out-of-plane geometry can have a reflection coefficient almost the same as in plane geometry [21]. The radiation polarization in the OC was discussed in [21, 23]. In [27] it was shown that in OC, effective smoothing of the output beam is possible if the angle $\gamma$ is different in the first and second pairs of gratings, which was later confirmed experimentally [28]. It was proposed to use the OC to increase the output power by reducing the incident angle $\alpha$ [29]. In a particular case of OC, when $\alpha$ is equal to the Littrow angle $\alpha_L$, the compressor "becomes" plane again, which greatly simplifies its experimental implementation. Such a compressor has a number of additional advantages [21], one of which is the possibility of using multilayer dielectric gratings, the reflection band of which rapidly narrows with increasing $(\alpha - \alpha_L)$, which makes them unsuitable for the TC in broadband lasers [26]. This compressor, given that it is symmetric, is called a Littrow compressor (LC). In general, an asymmetric, out-of-plane compressor (AOC) has 8 parameters; all the others are its particular cases (Table 1).

Both AC and OC are quite promising as they enable pulse power enhancement. However, the most important parameter is the focal intensity rather than the power. To increase focal intensity, the radiation should be Fourier-limited not only in time, but also in space, i.e. the pulse should have a constant spectral phase and the beam a constant spatial phase (plane wavefront). For this purpose, various dispersion management methods are widely used, including an acousto-optic programmable dispersive filter (AOPDF) [30], as well as adaptive mirrors [31]. Both these technologies are currently well developed and effectively correct temporal and spectral phase distortions separately, but, in principle, they cannot compensate for space-time coupling. Since diffraction gratings are not perfectly plane, any compressor inevitably introduces among others space-time coupling phase distortions, which reduce the focal intensity even with ideal operation of the AOPDF and adaptive mirror. This reduction has been numerically studied in many works [12, 32-37] for specific compressor parameters, but we are not aware of any analytical results. Different methods of space-time coupling compensation are also discussed in [12, 32-37], but all of them are difficult for experimental implementation.

In this work, the focal intensity is found analytically for an arbitrary shape of the surface of the diffraction gratings for the AOC, i.e. for any compressor from Table 1. In addition, a simple method of compensating for space-time coupling is proposed, which consists in two-angle adjusting of G4 grating, which allows us to significantly increase the focal intensity and/or reduce the requirements for the accuracy of grating manufacturing.

Table 1. Parameters of different types of compressors ($Y \equiv \alpha, N, \gamma, L$)

| Compressor | Plane (PC) $\gamma_1 = \gamma_2 = 0$ | Out-of-plane (OC) $\gamma_{1,2} \neq 0$ | Out-of-plane with $\alpha = \alpha_L$ $\gamma_{1,2} \neq 0$ |
|---|---|---|---|
| Symmetric (SC) $Y_{1,2} = Y_{1,2}$ | $\alpha$; $N$; $L$ Treacy compressor (TC) | $\alpha$; $N$; $\gamma$; $L$ | $N$; $\gamma$; $L$ Littrow compressor (LC) |
| Asymmetric (AC) $Y_{1,2} \neq Y_{1,2}$ | $\alpha_{1,2}$; $N_{1,2}$; $L_{1,2}$ | $\alpha_{1,2}$; $N_{1,2}$; $\gamma_{1,2}$; $L_{1,2}$ General case (AOC) | $N_{1,2}$; $\gamma_{1,2}$; $L_{1,2}$ |
| Two-grating $L_2 = 0$ | $\alpha$; $N$; $L$ | $\alpha$; $N$; $\gamma$; $L$ | $N$; $\gamma$; $L$ |

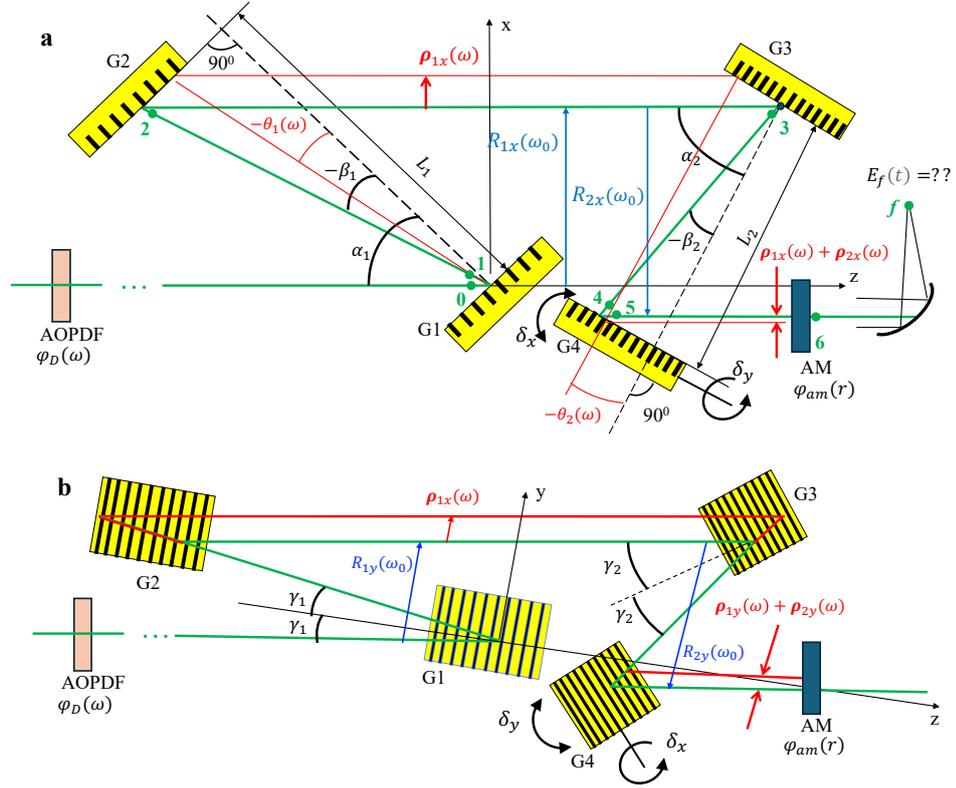

Fig. 1. Compressor schematic (asymmetric, out-of-plane); (a) top view, (b) side view. Green lines – beam at central frequency $\omega_0$, red lines – beam at arbitrary frequency $\omega$ shifted to the red part of the spectrum. AOPDF – acousto-optic programmable dispersive filter, AM – adaptive mirror.

## 2. Formulation of the problem

We will consider an AOC (Fig. 1) consisting of two pairs of gratings with line density $N_{1,2}$, distance between the gratings $L_{1,2}$ and angles of incidence on the first grating $\alpha_{1,2}$ and $\gamma_{1,2}$ in the zx and zy planes, respectively; the subscripts 1 and 2 correspond to the first and second grating pairs. All the other compressors listed in Table 1 are particular cases of the AOC. In the SC, the gratings G1 and G4 (as well as G2 and G3) are antiparallel in both planes, which corresponds to the replacement of the transverse wave vector $\boldsymbol{\kappa}$ by $(-\boldsymbol{\kappa})$ and, consequently, to the change in the sign of the angles of incidence: $\alpha_2 = -\alpha_1$ and $\gamma_2 = -\gamma_1$. For convenience, we will consider the second pair of gratings in a coordinate system rotated around the z axis by 180 degrees relative to the laboratory system for all angles of incidence $\alpha_{1,2}$ and $\gamma_{1,2}$ to be positive, i.e. $\alpha_2 = \alpha_1$ and $\gamma_2 = \gamma_1$ in the SC. Let the time spectrum of the field at each point of the cross section $\boldsymbol{r}$ be given at the input (point 0) in the form:

$$E_0(\omega, \boldsymbol{r}) = e^{i\varphi_{in}(\omega)+i\varphi_D(\omega)} \cdot e^{iHy} e^{i\varphi_0(\omega,\boldsymbol{r})} \cdot |E_0(\omega, \boldsymbol{r})|, \qquad (1)$$

where $\varphi_0(\omega, \boldsymbol{r})$ is the space phase characterizing the aberrations (wave front distortions) of the input field, $\varphi_{in}(\omega)$ is the spectrum phase without allowance for the phase introduced by the AOPDF $\varphi_D(\omega)$,

$$H = \frac{\omega}{c} \sin\gamma_1 \qquad (2)$$

The $(\omega, \boldsymbol{r})$-representation of the field is convenient for describing an imperfect grating surface with $h_n(\boldsymbol{r})$ profile, where $n = 1,2,3,4$ is the number of the grating. The incident and the reflected fields in this representation are related by the phase $\varphi_n(\omega, \boldsymbol{r})$:

$$E_{reflected}(\omega, \boldsymbol{r}) = e^{i\varphi_n(\omega,\boldsymbol{r})} E_{incident}(\omega, \boldsymbol{r}) \qquad (3)$$

In the model presented in the Appendix, the expression for $\varphi_n(\omega, \boldsymbol{r})$ is written as

$$\varphi_n(\omega, r) = -\frac{\omega}{c} d_n(\omega, r), \tag{4}$$

where

$$d_n(\omega, r) = \cos\gamma_{1,2}\left(\cos\theta_{1,2}(\omega) + \cos\alpha_{1,2}\right)h_n\left(\frac{x}{\cos\alpha_{1,2}}; \frac{y}{\cos\gamma_{1,2}}\right), \tag{5}$$

$$\sin\theta_{1,2}(\omega) = -\frac{2\pi c}{\omega \cos\gamma_{1,2}} N_{1,2} + \sin\alpha_{1,2} \tag{6}$$

Hereinafter the indices 1,2 for $\alpha, \theta, \gamma$ and $N$ correspond to the first ($n = 1,2$) and second ($n = 3,4$) grating pairs. Note that (4) is written in the reference frame aligned with the geometric center of the grating, rather than in the laboratory coordinate system. As can be seen from (5), after reflection from the grating, the phase front repeats the shape of the grating, but the proportionality coefficient depends on the frequency, which obviously leads to space-time coupling and a decrease in focal intensity. To the best of our knowledge, this effect has not been previously studied in the literature. In particular, an approximate model in which $d_n(\omega, r) = d_n(\omega_0, r)$ was considered in [12]. The field in the $(t, r)$-representation at any compressor point $j$ ($j=0,1…6$, see Fig. 1a) is an inverse Fourier transform of $E_j(\omega, r)$:

$$E_j(t, r) = \int E_j(\omega, r) e^{-i\omega t} d\omega \tag{7}$$

Here and below, all integrals are over infinite limits, and the multiplier $\sqrt{2\pi}$ in the Fourier transforms is omitted, as it does not affect the final result. Analogously, the field in the $(\omega, \kappa)$-representation is designated as $E_j(\omega, \kappa)$:

$$E_j(\omega, \kappa) = \int E_j(\omega, r) e^{i\kappa r} dr \tag{8}$$

It is convenient to describe the field propagation at reflection from two parallel gratings in the $(\omega, \kappa)$-representation, in which, for the minus first diffraction order, the input and output fields are related by the $\Psi_{p1,2}$ phase [10]:

$$\Psi_{p1,2}(\omega, k_x, k_y) = L_{1,2} k_{zx}\left(\cos\tilde{\theta}_{1,2} + \cos\left\{\alpha_{1,2} \pm \operatorname{atan}\frac{k_x}{k_z}\right\}\right), \tag{9}$$

where $k_{zx}^2 = \frac{\omega^2}{c^2} - k_y^2$, $k_z^2 = \frac{\omega^2}{c^2} - k_x^2 - k_y^2$, and $\tilde{\theta}$ is the angle of reflection from the grating:

$$\sin\tilde{\theta}_{1,2}(\omega, k_x, k_y) = -\frac{2\pi}{k_{zx}} N_{1,2} + \sin\left\{\alpha_{1,2} \pm \operatorname{atan}\frac{k_x}{k_z}\right\} \tag{10}$$

By neglecting the diffraction, (9) may be significantly simplified by expanding it into a Taylor series around the point $\begin{pmatrix} k_x \\ k_y \end{pmatrix} = \kappa_{1,2} = \begin{pmatrix} 0 \\ \frac{\omega}{c}\sin\gamma_{1,2} \end{pmatrix}$:

$$\Psi_{p1,2}(\omega, \kappa) = \Psi_{1,2}(\omega) + R_{1,2}(\omega)\kappa, \tag{11}$$

where

$$\Psi_{1,2}(\omega) = \Psi_{p1,2}(\omega, \kappa = \kappa_{1,2}) - R_{1,2}(\omega)\kappa_{1,2} \tag{12}$$

$$R_{1,2}(\omega) \equiv \frac{\partial \Psi_{p1,2}(\omega, \kappa = \kappa_{1,2})}{\partial \kappa} = \pm L \begin{pmatrix} -\frac{\sin(\theta_{1,2}(\omega) + \alpha_{1,2})}{\cos\theta_{1,2}(\omega)} \\ -\tan\gamma_{1,2} \frac{1 + \cos(\theta_{1,2}(\omega) + \alpha_{1,2})}{\cos\theta_{1,2}(\omega)} \end{pmatrix} \tag{13}$$

The $\pm$ sign in (9), (13) corresponds to the above convention on the positiveness of $\alpha_{1,2}$ and $\gamma_{1,2}$. The $-R_{1,2}(\omega)$ vector shows the beam displacement at frequency $\omega$ relative to the z axis. For $\alpha_1 > \alpha_{1L}(\omega)$ (the case depicted in Fig. 1a), $-R_{1x}(\omega) > 0$ for the first grating pair and $-R_{2x}(\omega) < 0$ for the second one. In the LC, $\alpha_1 = \alpha_{1L}(\omega_0)$; hence for $\omega < \omega_0$, $-R_{1x}(\omega) > 0$ and for $\omega < \omega_0$, $-R_{1x}(\omega) < 0$. In the SC, $R_1(\omega) + R_2(\omega) = 0$. In the PC, $R_{1y}(\omega) = R_{2y}(\omega) = 0$ – no beam displacement along the y axis. Note that, according to (6) and (10), $\theta_{1,2}(\omega) = \tilde{\theta}_{1,2}\left(\omega, k_x = 0, k_y = \frac{\omega}{c}\sin\gamma_{1,2}\right)$.

## 3. Focal intensity for arbitrary phase distortions $\varphi_n(\omega, r)$

Let us find the field at all points in Fig. 1a successively from the compressor input to the focus. According to (3),

$$E_1(\omega, r) = e^{i\varphi_1(\omega, r)} E_0(\omega, r), \tag{14}$$

from which, with allowance for (8),(9), we obtain

$$E_2(\omega, \kappa) = e^{i\Psi_{p1}(\omega, \kappa)} \int e^{i\varphi_1(\omega, r)} E_0(\omega, r) e^{i\kappa r} dr \tag{15}$$

Taking into account that the centers of the G2 and G3 gratings are at the point $r = R_1(\omega_0)$ in the laboratory frame of reference and using (3) we can find $E_3(\omega, r)$:

$$E_3(\omega, r) = e^{i\varphi_{23}(\omega, r - R_1(\omega_0))} E_2(\omega, r), \tag{16}$$

where $\varphi_{23} = \varphi_2 + \varphi_3$. Making the Fourier transform of (16), with (15), (14) and (11) taken into account, we find $E_3(\omega, \kappa)$:

$$E_3(\omega, \kappa) = e^{i\Psi_1(\omega)} e^{i\kappa R_1(\omega)} \int dr\, e^{i\varphi_{23}(\omega, r + \rho_1(\omega))} e^{i\kappa r} e^{i\varphi_1(\omega, r)} E_0(\omega, r), \tag{17}$$

where

$$\boldsymbol{\rho}_{1,2}(\omega) = \boldsymbol{R}_{1,2}(\omega) - \boldsymbol{R}_{1,2}(\omega_0) \tag{18}$$

is the beam displacement (difference of coordinates) at frequency $\omega$ relative to the beam at frequency $\omega_0$ (Fig. 1). The fields at the points 3 and 4 are related by the phase $\Psi_{p2}(\omega, \kappa)$:

$$E_4(\omega, \kappa) = e^{i\Psi_{p2}(\omega, \kappa)} E_3(\omega, \kappa) \tag{19}$$

The substitution of (17) into (19) taking into account (11) yields $E_4(\omega, \kappa)$, and after the inverse Fourier transform

$$E_4(\omega, r) = e^{i\Psi_2(\omega)} e^{i\Psi_1(\omega)} e^{i\varphi_{23}(\omega, r - R_{12}(\omega) + \rho(\omega))} e^{i\varphi_1(\omega, r - R_{12}(\omega))} E_0(\omega, r - R_{12}(\omega)), \tag{20}$$

where

$$\boldsymbol{R}_{12}(\omega) = \boldsymbol{R}_1(\omega) + \boldsymbol{R}_2(\omega) \tag{21}$$

From physical considerations, we can make an educated guess that the strongest space-time coupling distortions are nothing but an angular chirp. It can be compensated by adding a chirp of the same absolute value but of opposite sign, for which it is sufficient to simply rotate the G4 grating by angle $\delta_x$ in the xz plane and by angle $\delta_y$ in the yz plane (Fig. 1). With allowance for this and for the fact that the center of the G4 grating is located in the laboratory system at the point $r = R_{12}(\omega_0)$, we obtain

$$E_5(\omega, r) = E_4(\omega, r) e^{i\varphi_4(r - R_{12}(\omega_0))} e^{i\varepsilon(\omega)(r - R_{12}(\omega_0))}, \tag{22}$$

where $\boldsymbol{\varepsilon}(\omega)$ is the wave vector produced by the adjustment of the G4 grating. From (4) we find

$$\varepsilon_x = -\frac{\omega}{c}\frac{\cos\gamma_2}{\cos\alpha_2}(\cos\alpha_2 + \cos\theta_2(\omega))\delta_x \qquad \varepsilon_y = -\frac{\omega}{c}(\cos\theta_2(\omega) + \cos\alpha_2)\delta_y \tag{23}$$

The adaptive mirror introduces the phase $\varphi_{am} = -2\frac{\omega}{c}h_{am}(r)$, where $h_{am}(r)$ is its surface shape. The mirror center is at the point $r = R_{12}(\omega_0)$, therefore

$$E_6(\omega, r) = E_5(\omega, r) e^{i\varphi_{am}(\omega, r - R_{12}(\omega_0))} \tag{24}$$

By substituting (20) into (22) and the result into (24), we find $E_6(\omega, \mathbf{r})$ and after the Fourier transform $E_6(\omega, \boldsymbol{\kappa})$:

$$E_6(\omega, \boldsymbol{\kappa}) = e^{i\Psi_{12}(\omega)} e^{i\varphi_{in}(\omega)+i\varphi_D(\omega)} e^{i\boldsymbol{\kappa}\mathbf{R}_{12}(\omega)} \int d\mathbf{r}\, e^{i\boldsymbol{\kappa}\mathbf{r}} e^{i\boldsymbol{\varepsilon}(\omega)\mathbf{r}} e^{iHy} e^{i\phi(\omega,\mathbf{r})} |E_0(\omega,\mathbf{r})|, \quad (25)$$

where

$$\Psi_{12}(\omega) = \Psi_1(\omega) + \Psi_2(\omega) + \boldsymbol{\varepsilon}(\omega)\big(\mathbf{R}_{12}(\omega) - \mathbf{R}_{12}(\omega_0)\big), \quad (26)$$

$$\phi(\omega, \mathbf{r}) = \varphi_0(\omega, \mathbf{r}) + \varphi_1(\omega, \mathbf{r}) + \varphi_{23}(\omega, \mathbf{r} + \boldsymbol{\rho}_1(\omega)) + \varphi_4(\omega, \mathbf{r} + \boldsymbol{\rho}_1(\omega) + \boldsymbol{\rho}_2(\omega)) + \varphi_{am}(\omega, \mathbf{r} + \boldsymbol{\rho}_1(\omega) + \boldsymbol{\rho}_2(\omega)) \quad (27)$$

From (27) it is clear that the adaptive mirror most effectively (although not completely) compensates for the distortions $\varphi_4$ of the G4 grating, since the spatial chirp on the mirror and on the grating is the same. In the SC, $\boldsymbol{\rho}_1(\omega) + \boldsymbol{\rho}_2(\omega) = 0$ (see (13), (18)), the input beam distortions $\varphi_0$ are completely compensated for, and the distortions $\varphi_1$ of the G1 grating are compensated for just as well as $\varphi_4$. This is not the case in the AC. However, if the asymmetry is not large, i.e. $\boldsymbol{\rho}_1(\omega) + \boldsymbol{\rho}_2(\omega) \ll \boldsymbol{\rho}_{1,2}(\omega)$, then the differences from the SC may be neglected. In a strongly asymmetric compressor (for example, a double-grating compressor), an additional adaptive mirror located in front of the compressor should be used to completely compensate for $\varphi_0$ and to a large extent for $\varphi_1$. This reduces the problem to the case of a SC, so we will further consider this particular case. Then (25) can be reduced to the form

$$E_6(\omega, \boldsymbol{\kappa}) = e^{i\Psi_{12}(\omega)} e^{i\varphi_{in}(\omega)+i\varphi_D(\omega)} e^{i\Psi_{aber}(\omega)} e^{i\boldsymbol{\kappa}\mathbf{R}_{12}(\omega)} \int d\mathbf{r}\, e^{i\boldsymbol{\kappa}\mathbf{r}} e^{i\boldsymbol{\varepsilon}(\omega)\mathbf{r}} e^{iHy} e^{i\phi_{23}(\omega,\mathbf{r})+i\phi_{all}(\omega,\mathbf{r})} |E_0(\omega,\mathbf{r})|, \quad (28)$$

where

$$\phi_{23}(\omega, \mathbf{r}) = -\frac{\omega}{c}\sum_{j=2}^{3}\big(d_j(\omega_0, \mathbf{r} + \boldsymbol{\rho}_1(\omega)) - d_j(\omega_0, \mathbf{r} + \boldsymbol{\rho}_1(\omega) + \boldsymbol{\rho}_2(\omega)) - d_j(\omega_0, \boldsymbol{\rho}_1(\omega)) + d_j(\omega_0, \boldsymbol{\rho}_1(\omega) + \boldsymbol{\rho}_2(\omega))\big) \quad (29)$$

$$\phi_{all}(\omega, \mathbf{r}) = -\frac{\omega}{c}\sum_{j=1}^{4}\big(d_j(\omega, \mathbf{r}) - d_j(\omega_0, \mathbf{r})\big) \quad (30)$$

$$\Psi_{aber}(\omega) = -\frac{\omega}{c}\sum_{j=2}^{3}\big(d_j(\omega_0, \boldsymbol{\rho}_1(\omega)) - d_j(\omega_0, \boldsymbol{\rho}_1(\omega) + \boldsymbol{\rho}_2(\omega))\big) \quad (31)$$

In (28) we zeroed the sum of all terms of the phase of the form $\omega f(\mathbf{r})$, i.e. we zeroed the phase that may be compensated by an adaptive mirror and obtained for the shape of the mirror surface the following condition

$$2h_{am}(\mathbf{r}) = -\sum_{j=0}^{4} d_j(\omega_0, \mathbf{r}) - \frac{c}{\omega}\varphi_0(\omega, \mathbf{r}), \quad (32)$$

i.e. the surface of the adaptive mirror (except for the compensation of the distortions of the input wave front $\frac{c}{\omega}\varphi_0(\omega, \mathbf{r})$) will repeat with inverse sign the total shape of all wave front distortions for the radiation at frequency $\omega_0$, rather than the total shape of all surfaces $\sum_{j=0}^{4} h_j(\mathbf{r})$. In other words, the adaptive mirror must optimize the radiation wave front at frequency $\omega_0$. Next, we will find the field at the focal point, i.e. at $Hy + (\boldsymbol{\kappa} + \boldsymbol{\varepsilon}(\omega_0))\mathbf{r} = 0$:

$$E_f(\omega) = E_6\big(\omega, \kappa_x = -\varepsilon_x(\omega_0), \kappa_y = -\varepsilon_y(\omega_0) - H\big) \quad (33)$$

Substituting (25) into (33) and passing over to the frequency $\Omega = \omega - \omega_0$, yields

$$E_f(\Omega) = \int d\mathbf{r}\, e^{i\boldsymbol{\mu}(\Omega)\mathbf{r}} e^{i\phi_{23}(\Omega,\mathbf{r})+i\phi_{all}(\Omega,\mathbf{r})} |E_0(\Omega, \mathbf{r})|, \quad (34)$$

where

$$\boldsymbol{\mu}(\Omega) = \boldsymbol{\varepsilon}(\Omega) - \boldsymbol{\varepsilon}(\Omega_0) = \frac{\Omega}{\omega_0}\mathbf{A}_\mu + C_\mu\left(\frac{\Omega}{\omega_0}\right)^2 \mathbf{A}_\mu + O\left(\left(\frac{\Omega}{\omega_0}\right)^3\right), \quad (35)$$

$$A_\mu = -k_0 w \frac{1+\cos(\alpha_2+\beta_2)}{\cos\alpha_2 \cos\beta_2}\begin{pmatrix}\delta_x \cos\gamma_2 \\ \delta_y \cos\alpha_2\end{pmatrix} \quad C_\mu = -\frac{(\sin\alpha_2-\sin\beta_2)^2}{2\cos^2\beta_2(1+\cos(\alpha_2+\beta_2))}, \qquad (36)$$

and $\beta_{1,2} = \theta_{1,2}(\omega_0)$ is the angle of refection from the grating at frequency $\omega_0$ (Fig. 1a). In (34) we zeroed the sum of all phase terms dependent only on $\Omega$, i.e. we zeroed the phase that may be compensated for by AOPDF and obtained for the AOPDF-introduced phase $\Psi_D(\Omega)$ the condition

$$\varphi_D(\Omega) = -\varphi_{in}(\Omega) - \Psi_1(\Omega) - \Psi_2(\Omega) + HR_{12y}(\Omega) - \Psi_{aber}(\Omega) + \boldsymbol{\mu}(\Omega)(\boldsymbol{\rho}_1(\Omega) + \boldsymbol{\rho}_2(\Omega)) \qquad (37)$$

The first four terms correspond to a compressor without space-time coupling (for a PC, only the first three terms remain). The fifth term generalizes the result to the case of space-time coupling. It permits obtaining the shortest possible pulse exactly at the focal point, rather than in the near field. The spectral phases of these two pulses differ by $\Psi_{aber}(\omega)$ that can be easily found from (5), (31), if surface shapes of G2 and G3 gratings are known. Thus, dispersion management significantly depends on space-time coupling. Finally, the last term with $\boldsymbol{\mu}(\omega)$ takes into account the correction of the angle of the fourth grating (note that in the SC this term is equal to zero).

From (34) it is clearly seen that space-time coupling involves two effects described by the phase terms $\phi_{23}(\Omega, \boldsymbol{r})$ (29) and $\phi_{all}(\Omega, \boldsymbol{r})$ (30). The first effect is due to the fact that different frequencies are reflected from different areas of G2 and G3. G1 and G4 gratings do not contribute to this effect, as it is completely compensated by the adaptive mirror (see above). The second effect, on the contrary, occurs in all four gratings and is associated with the fact that, since the spatial phase of the beam reflected from the grating does not exactly follow the shape of the grating surface (see (4)). As far as we know, this effect has not been studied in the literature before. In particular, only the first effect was taken into account in [12, 32-37]. The term $\boldsymbol{\mu}(\omega)\boldsymbol{r}$ in the exponent in (34) is determined by the detuning of two angles of incidence on G4 grating, $\delta_x$ and $\delta_y$, and depends strictly linearly on $\boldsymbol{r}$. This means that only a linear angular chirp (the linear dependence of the wave vector direction on the frequency) can be compensated.

From (29) and (30) it is clear that $\phi_{23}(\Omega = 0, \boldsymbol{r}) = \phi_{all}(\Omega = 0, \boldsymbol{r}) = 0$. In the first order, of both $\phi_{23}(\Omega, \boldsymbol{r})$ and $\phi_{all}(\Omega, \boldsymbol{r})$ are proportional to the first power of $\Delta = \frac{\Delta\omega}{\omega_0} \ll 1$. Besides, they are also proportional to the first power of the grating surface shape and, therefore, are proportional to the rms of the surface shape σ, i.e. the sum $\phi_{23}(\Omega, \boldsymbol{r}) + \phi_{all}(\Omega, \boldsymbol{r})$ is proportional to σΔω. Thus, from (29), (30) and (34) it follows that, from the point of view of space-time coupling effects, a decrease in Δω is equivalent to the corresponding decrease in σ and vice versa. Consequently, with an increase in Δω, the behavior of the intensity at the focal point is contradictory: it increases in proportion to 1/Δω due to the shortening of the Fourier-limited pulse, and decreases due to space-time coupling, thus moving the pulse away from the Fourier limit.

The field in focus $E_f(t)$ is the inverse Fourier transform of (34) in time. In what follows, we will assume that the input field is super-Gaussian both in time and in both spatial coordinates (such beams are typical for high-power lasers, especially OPCPA lasers [12, 38-40]):

$$|E_0(\Omega, x, y)| = |E_{00}|e^{-\xi^{2\mu}}e^{-x^{2\nu}}e^{-y^{2\nu}}, \qquad (38)$$

where $\xi = \frac{\Omega}{\Delta\omega}$, $\boldsymbol{u} = (x, y) = \frac{\boldsymbol{r}}{w}$ and $T = t\Delta\omega$ are the dimensionless frequency, coordinates and time, Δω is the spectrum halfwidth at the 1/e level of the field, and $w$ is the beam halfwidth at the $1/e$ level of the field. From (34), (38) we obtain an expression for the pulse shape at the focal point:

$$E_f(T) = |E_{00}|\Delta\omega w^2 \int d\xi e^{-i\xi T}e^{-\xi^{2\mu}} \int d\boldsymbol{u} e^{-x^{2\nu}}e^{-y^{2\nu}} e^{i(\boldsymbol{\mu}(\xi)\boldsymbol{u}+\phi_{23}(\xi,\boldsymbol{u})+\phi_{all}(\xi,\boldsymbol{u}))} \qquad (39)$$

By zeroing the time derivative of (39) we can find the time $T_{max}$, at which $E_f(T)$ is maximal and the maximum focal intensity is $|E_f(T = T_{max})|^2$, from which follows the Strehl ratio for the focal intensity

$$St = \frac{|E_f(T_{max})|^2}{|E_f(T=0;\boldsymbol{\mu}=\phi_{23}=\phi_{all}=0)|^2}, \qquad (40)$$

which shows the focal intensity decrease caused by the space-time coupling. From (39) it is clear that $T_{max} = 0$ in the absence of space time coupling, i.e. if $\boldsymbol{\mu} = \phi_{23} = \phi_{all} = 0$. Since in the described model the temporal distortions are fully compensated by AOPDF and the spatial distortions by the adaptive mirror, in the absence of space-time coupling $St = 1$.

## 4. Grating surface in the form of quadratic Zernike polynomials

Without loss of generality, wedge-shaped aberrations may be regarded to be equal to zero. Let the grating surface shape $h_n(\boldsymbol{r})$ be a sum of three quadratic Zernike polynomials – defocus $z_2^0(\boldsymbol{r})$, vertical astigmatism $z_2^2(\boldsymbol{r})$ and oblique astigmatism $z_2^{-2}(\boldsymbol{r})$ with Zernike coefficients $Z^0, Z^2$ and $Z^{-2}$:

$$h_n(\boldsymbol{r}) = \frac{c}{\omega_0}(Z_n^0 z_2^0(\boldsymbol{r}) + Z_n^2 z_2^2(\boldsymbol{r}) + Z_n^{-2} z_2^{-2}(\boldsymbol{r})) \qquad (41)$$

Then, (29) and (30) transform to

$$\phi_{23}(\xi, \boldsymbol{u}) = \left(\xi\Delta\frac{L_2}{w}\boldsymbol{q} + (\xi\Delta)^2\frac{L_2}{w}\boldsymbol{q_2}\right)\boldsymbol{u} + O\left(\frac{L_2}{w}\Delta^3\right), \qquad (42)$$

$$\phi_{all}(\omega, x, y) = \xi\Delta(Px^2 + Qy^2 + Sxy) + O(\Delta^2), \qquad (43)$$

where $\boldsymbol{q}, \boldsymbol{q_2}, P, Q,$ and $S$ depend only on the compressor geometric parameters $\alpha_{1,2}, \beta_{1,2}, \gamma_{1,2}$ and on Zernike coefficients:

$$\boldsymbol{q} = (\hat{\boldsymbol{Z}}_2 + \hat{\boldsymbol{Z}}_3)\boldsymbol{A} \qquad \boldsymbol{q_2} = (\hat{\boldsymbol{Z}}_2 + \hat{\boldsymbol{Z}}_3)(\boldsymbol{A} - \boldsymbol{B}) \qquad (44)$$

$$\boldsymbol{A} = \begin{pmatrix} \frac{\cos\alpha_2}{\cos^3\beta_2}(\sin\alpha_2 - \sin\beta_2) \\ \tan\gamma_2 \frac{(\sin\alpha_2 - \sin\beta_2)^2}{\cos^3\beta_2} \end{pmatrix} \quad \boldsymbol{B} = \begin{pmatrix} \frac{\cos\alpha_2}{\cos^3\beta_2}(\sin\alpha_2 - \sin\beta_2)\left(1 - \frac{3}{2}\sin\beta_2\frac{\sin\alpha_2-\sin\beta_2}{\cos^2\beta_2}\right) \\ \tan\gamma_2 \frac{(\sin\alpha_2 - si\ \beta_2)^2}{\cos^3\beta_2}\left(1 + \frac{\cos\beta_2 + 3(\sin\alpha_2 - \sin\beta_2)}{2\cos^2\beta_2}\right) \end{pmatrix} \qquad (45)$$

$$\hat{\boldsymbol{Z}}_n = \cos\gamma_{1,2}\left(\cos\beta_{1,2} + \cos\alpha_{1,2}\right)\begin{pmatrix} \frac{2(Z_n^0 2\sqrt{3} + Z_n^2\sqrt{6})}{\cos\alpha_{1,2}} & \frac{\sqrt{6}Z_n^{-2}}{\cos\alpha_{1,2}} \\ \frac{\sqrt{6}Z_n^{-2}}{\cos\gamma_{1,2}} & \frac{2(Z_n^0 2\sqrt{3} - Z_n^2\sqrt{6})}{\cos\gamma_{1,2}} \end{pmatrix} \qquad (46)$$

$$P = \sum_{n=1}^{4} p_n \qquad Q = \sum_{n=1}^{4} q_n \qquad S = \sum_{n=1}^{4} s_n \qquad (47)$$

$$\begin{pmatrix} p_n \\ q_n \\ s_n \end{pmatrix} = \cos\gamma_{1,2}\, tg\beta_{1,2}(\sin\alpha_{1,2} - \sin\beta_{1,2})\sqrt{6}\begin{pmatrix} \frac{1}{\cos^2\alpha_{1,2}}(Z_n^0\sqrt{2} + Z_n^2) \\ \frac{1}{\cos^2\gamma_{1,2}}(Z_n^0\sqrt{2} - Z_n^2) \\ \frac{1}{\cos\alpha_{1,2}\cos\gamma_{1,2}}Z_n^{-2} \end{pmatrix} \qquad (48)$$

### 4.1. Without compensation of space-time coupling

Without compensation, $\boldsymbol{\mu}(\Omega) = 0$. In this case, by substituting (43), (42) into (39) and neglecting the terms of order $\Delta^2\frac{L_2}{w}$ and $\Delta$ in the exponent we obtain

$$E_f(T) = |E_{00}|\Delta\omega w^2 \int d\xi e^{-i\xi T} e^{-\xi^{2\mu}} \int d\boldsymbol{u}\, e^{-x^{2\nu}} e^{-y^{2\nu}} \cos\left(\xi\Delta\frac{L_2}{w}\boldsymbol{q}\boldsymbol{u}\right), \qquad (49)$$

From which follows that $T_{max} = 0$. If $\xi\Delta\frac{L_2}{w}q_{1x,1y} \ll 1$, from (49) and (40) we find

$$St = 1 - \frac{\Gamma\left(\frac{3}{2\nu}\right)\Gamma\left(\frac{3}{2\mu}\right)}{\Gamma\left(\frac{1}{2\nu}\right)\Gamma\left(\frac{1}{2\mu}\right)}\Delta^2\left(\frac{L_2}{w}\right)^2 \boldsymbol{q}^2, \qquad (50)$$

where $\Gamma$ is a Gamma function. From (49) it is clear that the space-time coupling is a linear angular chirp – the effective wave vector $\xi\Delta\frac{L_2}{w}\boldsymbol{q}$ is proportional to the frequency $\Omega$. As was to be expected, the other effects are much weaker, so their impact on $St$ may be neglected. It is important to note that both $E_f(T)$ and $St$ depend on the surface shape of the gratings G2 and G3 and do not depend on G1 and G4 (see (44)). To be more exact, the dependence on G1 and G4 is so weak that it may be neglected. In addition, $E_f(T)$ and $St$ depend on the geometric parameters $\alpha_2, \beta_2, \gamma_2, L_2$ of the second pair of gratings and do not depend on the parameters of the first pair. Physically, this is explained by the fact that of major importance is the difference between the spatial chirps on the grating and on the adaptive mirror, rather than the absolute value of the spatial chirp on the gratings.

## 4.2. Compensating for space-time coupling by rotating the fourth grating

From (42) and (35) it follows that, in the last exponent in (39), the largest term proportional to $\Delta\frac{L_2}{w}$ turns to zero if

$$\boldsymbol{A}_\mu = -\frac{L_2}{w}\boldsymbol{q} \tag{51}$$

Physically, this means that the linear angular chirp resulting from the imperfect surfaces of G2 and G3 is completely compensated by adjusting the angle of the G4 grating. Note that $\boldsymbol{A}_\mu$ and $\boldsymbol{q}$ are vectors, i.e. the chirp must be compensated in both planes. From (36) and (51) we obtain the adjustment angles $\delta_{x,y}$ for G4:

$$\begin{pmatrix}\delta_x\\\delta_y\end{pmatrix} = \frac{1}{k_0 w}\frac{L_2}{w}\frac{\cos\alpha_2\cos\beta_2}{1+\cos(\alpha_2+\beta_2)}\begin{pmatrix}\frac{q_x}{\cos\gamma_2}\\\frac{q_y}{\cos\alpha_2}\end{pmatrix} \tag{52}$$

The substitution of (42), (35), (51) into (39) yields

$$E_f(T) = |E_{00}|\Delta\omega w^2 \int d\xi\, e^{-\xi^{2\mu}} \int d\boldsymbol{u}\, e^{-x^{2\nu}}e^{-y^{2\nu}} \cos\left(\xi\Delta(Px^2+Qy^2+Sxy) + (\xi\Delta)^2\frac{L_2}{w}\boldsymbol{s}\boldsymbol{u} - \xi T\right) \tag{53}$$

where $\boldsymbol{s} = \boldsymbol{q}_2 - C_\mu \boldsymbol{q}$. If the cosine expression is much less than 1, then from (53), (40) we obtain the Strehl ratio in the case of compensation $St_{comp}$:

$$St_{comp} = 1 - G_{23} - G_{all}, \tag{54}$$

where

$$G_{23} = \Delta^4\left(\frac{L_2}{w}\right)^2 \frac{\Gamma\left(\frac{5}{2\mu}\right)\Gamma\left(\frac{3}{2\nu}\right)}{\Gamma\left(\frac{1}{2\mu}\right)\Gamma\left(\frac{1}{2\nu}\right)}\boldsymbol{s}^2 \tag{55}$$

$$G_{all} = \Delta^2 \frac{\Gamma\left(\frac{3}{2\mu}\right)}{\Gamma^2\left(\frac{1}{2\mu}\right)\Gamma^2\left(\frac{1}{2\nu}\right)}\left\{(P^2+Q^2)\left(\Gamma\left(\frac{1}{2\mu}\right)\Gamma\left(\frac{5}{2\nu}\right)\Gamma\left(\frac{1}{2\nu}\right) - \Gamma\left(\frac{3}{2\mu}\right)\Gamma^2\left(\frac{3}{2\nu}\right)\right) + S^2\Gamma\left(\frac{1}{2\mu}\right)\Gamma^2\left(\frac{3}{2\nu}\right)\right\} \tag{56}$$

The comparison of (50) and (54) shows that the difference of $St_{comp}$ from unity is much less than the difference of $St$ from unity, as $\Delta \ll 1$ and $w \ll L_2$. $St_{comp}$ decreases due to two effects. The first effect is the spatial chirp on the gratings G2 and G3 (the term $G_{23}$). As a result of the adjustment of G4 in (54) this term is proportional to $\Delta^4$, rather than to $\Delta^2$ like in (50). The second effect is associated with the fact that, although the spatial phase of the beam reflected from the grating is proportional to the grating surface shape, the proportionality coefficient depends on frequency (see (4), and all four gratings contribute to it (the term $G_{all}$). Since $\Delta \ll 1$ and a $L_2 \gg w$, the relationship between these two terms can be arbitrary and, in a general case, we cannot neglect any of them.

Next, let us set, as an example, four options for distortions of the surface shape of the gratings: only defocus, only vertical astigmatism, only oblique astigmatism, and the sum of these three polynomials with coefficients related as $Z_n^0: Z_n^2: Z_n^{-2} = 1: \frac{\sqrt{2}}{3}: \frac{\sqrt{2}}{3}$, which corresponds to the astigmatism $1 \pm 1/3$. It is convenient to express the answer in terms of the rms of the average surface shape $\frac{1}{4}\sum_{n=1}^{4} h_n(r)$, i.e. the rms of the surface of one (average) grating:

$$\sigma^2 = \frac{1}{w^2} \int_{-w/2}^{w/2} dx \int_{-w/2}^{w/2} \left(\frac{1}{4}\sum_{n=1}^{4} h_n(r)\right)^2 dy \qquad (57)$$

Note that the rms is calculated over a square with side $w$, rather than over the entire grating area.

### 4.3. Plane symmetric compressor (Treacy compressor)

Consider as an example the TC parameters of XCELS from [29]: $L_2 = 7.67w$, $\Delta = 0.0824$, $\alpha_2 = 36°$, $\gamma_2 = 0$, $N=950$/mm, $\nu = 6$, and $\mu = 6$. As mentioned above, as a result of space-time coupling the spectrum phases of a pulse in the near field and in focus differ by $\Psi_{aber}(\omega)$ that can be readily found from (5), (31), given known surface shapes of the gratings G2 and G3. The dependence of $GVD = \left|\frac{1}{2}\frac{d^2}{d\omega^2}\Psi_{aber}\right|$ on $\sigma$ is shown in Fig. 2 a. Since the duration of the Fourier-limited pulse is 17 fs, the neglect of $\Psi_{aber}(\omega)$ will lead to a significant lengthening of the pulse in focus. In practice, this means that the minimum pulse width in the near field cannot serve as a feedback for generating the AOPDF signal, so the pulse width at the focal point must be used.

The curves for $\delta_{x,y}(\sigma)$ are plotted using formula (52) in Fig. 3a. The defocus and the vertical astigmatism (with equal σ) have equal values of $\delta_x$, and their $\delta_y$ are equal in magnitude but opposite in sign. Therefore, the plots include data for three variants of the surface shape: defocus, oblique astigmatism and the sum of three polynomials. The graphs are plotted assuming that the Zernike coefficients are positive. Otherwise, $\delta_{x,y}(\sigma)$ changes its sign. As can be seen from the figure, in the last two cases $\delta_y \neq 0$, i.e. even for the TC, the G4 grating should be adjusted not only in the horizontal, but also in the vertical plane (out of diffraction plane). Note that $\delta_{x,y}$ can significantly exceed the diffraction angle that is about 0.3 arcseconds.

The $St$ values for defocus and direct astigmatism (with equal σ) are identical. The curves for $St(\sigma)$ (dashed curves), as well as for $(1 - G_{all})$ (dash-dot curves) and $(1 - G_{23})$ (dotted curves) for comparison of the contributions from different effects are plotted in Fig. 4a using the approximate formulas (50, 54). The solid curves are plotted using the exact formulas (49), (53). It can be seen from the figure that the approximate formulas (50, 54) give a very accurate result for $St > 0.75$. Without compensation, $St \ll 1$, even at $\sigma = \lambda/12$, i.e. in practice it is hard to obtain $St \approx 1$. However, in the case of compensation, the requirements for the surface quality of the gratings are quite realistic: $St > 0.83$ for $\sigma = \lambda/2$.

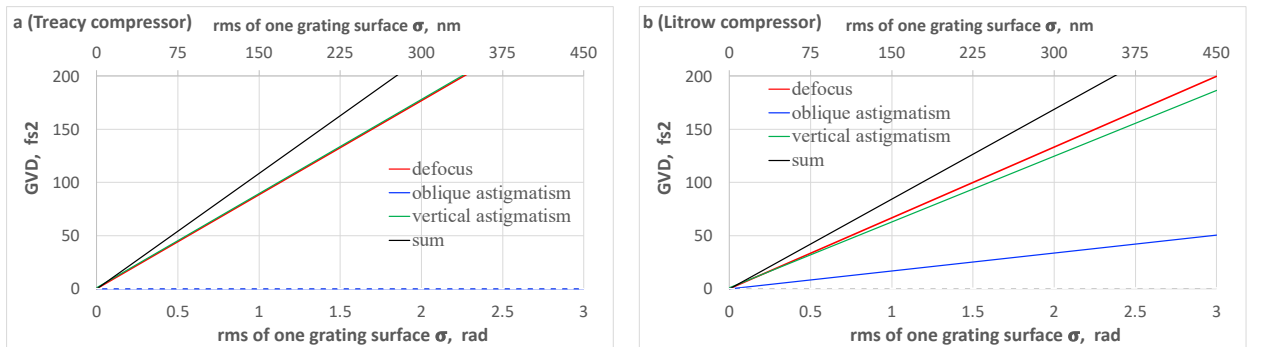

Fig. 2. (a) Treacy compressor, (b) Littrow compressor. The difference of GVD pulses in the near field and at the focal point: defocus (blue), or vertical astigmatism (green), oblique astigmatism

(red) and the sum of three Zernike polynomials (black). The curves for defocus and direct astigmatism in Fig. 2a coincide.

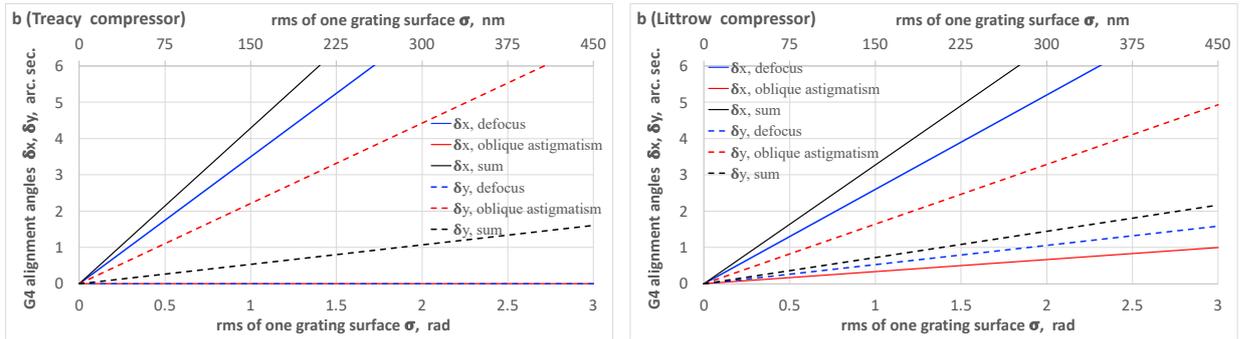

Fig. 3. (a) Treacy compressor, (b) Littrow compressor. Adjustment angle of G4 grating in the x plane $\delta_x$ (solid curves) and in the y plane $\delta_y$ (dashed curves) for the grating shape shaped as defocus or vertical astigmatism (blue), oblique astigmatism (red) and the sum of three Zernike polynomials (black). In Fig. 3a, $\delta_x$ for oblique astigmatism and $\delta_y$ for defocus are equal to zero.

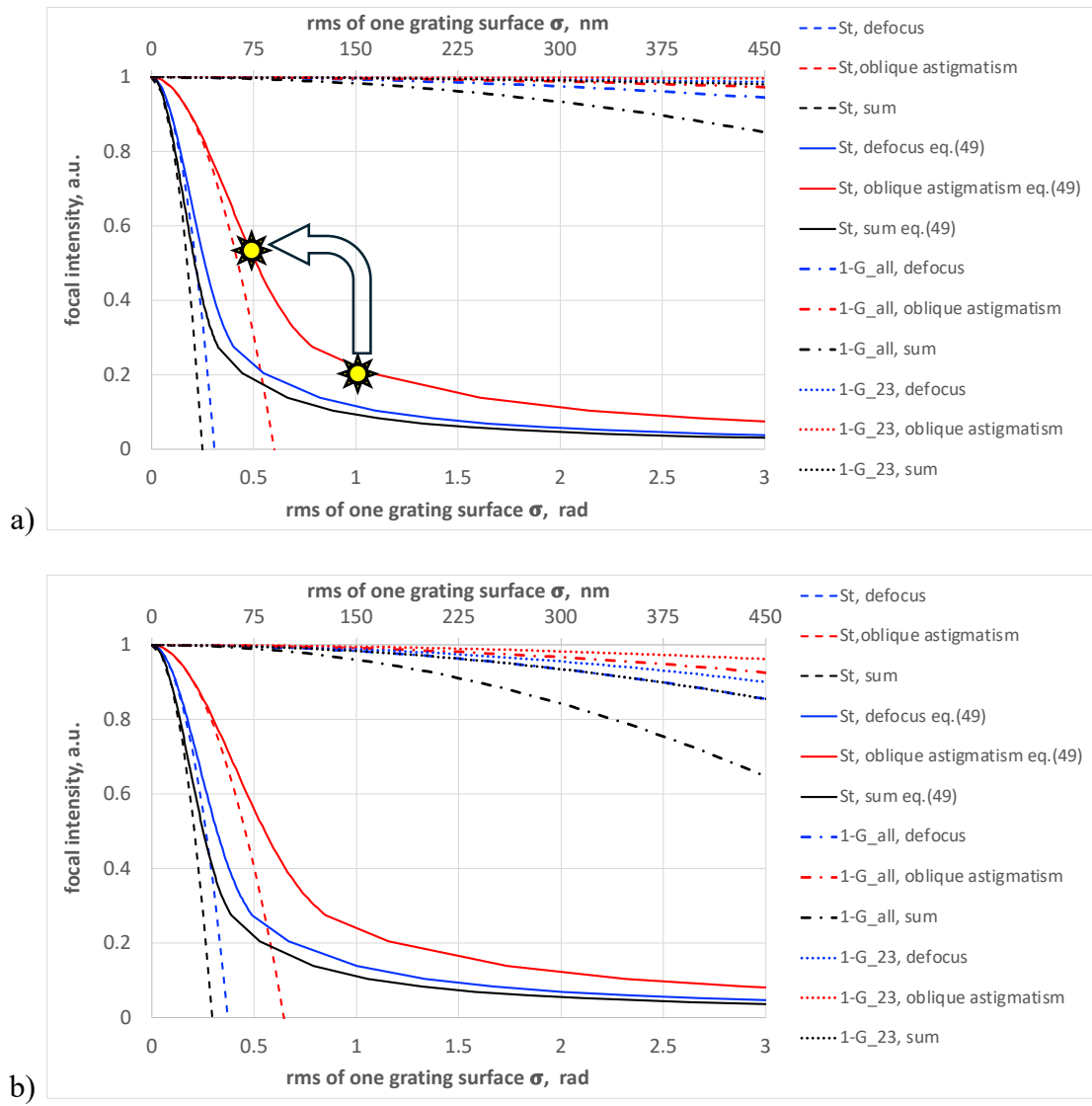

Fig. 4. (a) Treacy compressor, (b) Littrow compressor. Focal intensity (Strehl ratio) as a function of the rms of one grating surface for the grating shaped as defocus or vertical astigmatism (blue), oblique astigmatism (red) and the sum of three Zernike polynomials (black), plotted by approximate

formulas: without compensation (50) (dashed curves) and with compensation (54) (dash-dotted curves correspond to $1 - G_{all}$, and dotted cures to $1 - G_{23}$). The solid curves are plotted by the exact formula (49).

### 4.4. Out-of-plane symmetric compressor (Littrow compressor)

Consider as an example the LC parameters of XCELS from [29]: $L_2 = 4.8w$, $\Delta = 0.0824$, $\alpha_2 = 27.4°$, $\gamma_2 = 11.2°$, $N$=1000/mm; $\nu = 6$; $\mu = 6$. Analogously to Figs. 2-4a, plots for $GVD(\sigma)$ are shown in Fig. 2b, for $\delta_{x,y}(\sigma)$ in Fig. 3b, and for the Strehl ratio in Fig. 4b. As seen from Fig. 4b, formulas (50, 54) give a very accurate result at $St > 0.75$. Without compensation, the results almost coincide with those for the TC: $St \ll 1$, even for $\sigma = \lambda/12$. In the case of compensation, the requirements for the quality of grating surface are much weaker but more stringent than for the TC: $St > 0.82$ for $\sigma = \lambda/3$.

In practice, it is reasonable to measure the surface shape of each grating and then select the most successful grating sequence. If for some reason compensation is impossible, then it is necessary to minimize $\boldsymbol{q}^2$ the contribution to which is made only by the G2 and G3 gratings, see (44). Thus, from four gratings it is necessary to select a pair for which the algebraic sum of surface shapes gives the minimum value of $\boldsymbol{q}^2$. In the case with compensation, it is necessary to choose a sequence of four gratings that gives the maximum value of the Strehl ratio calculated by formula (54).

### 4.5. Asymmetric compressor

Without compensation, the asymmetric compressor is almost like the symmetric one, as according to (50) $St$ depends on the geometric parameters $\alpha_2, \beta_2, \gamma_2, L_2$ of the second pair of gratings only. Therefore, to increase $St$ it is necessary to choose $L_2 < L_1$. In particular, for a two-grating compressor, $\boldsymbol{q} = 0$, as $L_2 = 0$. This means that, for a two-grating compressor, no compensation is required and $St = St_{comp}$. Physically, this is explained by the fact that the spatial chirp on the adaptive mirror is exactly the same as on the G2 grating.

## 5. Arbitrary grating surface

In Section 4, we assumed only three quadratic Zernike polynomials to be nonzero. In a general case, there are also other polynomials with a power higher than 2. Then in expressions (43), (42) additional terms with higher-order x and y will appear, but the condition for optimal compensation (51) will remain the same. Thus, in a general case, it is necessary to choose a sequence of gratings that will give the maximum value of (39), (40); note that, when calculating $\boldsymbol{\mu}$ in (39), (52) shall be used. It should be taken into account that each grating may be rotated by 180 degrees around the normal by reversing the top and bottom, which will change the sign of all odd Zernike polynomials. Consequently, in addition to the sequence of gratings, it is also necessary to choose their optimal mutual orientation.

Note that the conclusion made in Section 4 about almost complete compensation of the influence of spatial chirp on gratings G2 and G3 is not correct in a general case. From (40) it can be shown that, for aberrations in the form of a coma, the contribution of the G2 and G3 gratings to the reduction of $St$ is dominant. In this case, space-time coupling gives rise to the effects at the compressor output that cannot be compensated by rotating G4, for example, defocus chirp, i.e. frequency dependent defocus. From the four gratings it is necessary to select two (taking into account the possibility of rotation by 180 degrees) that would give the smallest value of total aberrations after subtracting three quadratic Zernike polynomials. These gratings will be used as G2 and G3. The location and orientation of the remaining two gratings shall be chosen taking into account the maximization of (39). If the quality of the gratings is such that the value of $St$ remains unacceptably small, then the defocus chirp can be compensated by placing a lens telescope between the stretcher and the compressor with a defocus chirp of the same magnitude but opposite sign, or using the defocus chirp introduced by the stretcher.

As stated above, a decrease in $\Delta\omega$ is equivalent to the corresponding decrease in $\sigma$. A decrease in $\Delta\omega$ by a factor of 2 will lead to a twofold decrease in power, but, as shown by the arrow in Fig. 3a, if $St \approx 0.2$, this will simultaneously lead to an increase in $St$ by approximately 2.3 times. Consequently, the focal intensity will not decrease; on the contrary, it will increase. At the same time, the 2-fold decrease in $\Delta\omega$ makes it possible to reduce the size of the diffraction gratings, as well as significantly simplify the design of the laser as a whole. This circumstance may be useful when designing high-power lasers.

## 6. Conclusions and further research

Basic results of the study:

1. An analytical expression for the focal intensity of a laser pulse as a function of the surface shape of compressor diffraction gratings has been obtained. The expression has been obtained in a general form for an arbitrary compressor – an asymmetric out-of-plane compressor (see Table 1).

2. The focal intensity decreases as a result of an imperfect shape of the gratings caused by two effects. The first of them is the spatial chirp of the beam on the G2 and G3 gratings, due to which different frequencies face different surface shapes. The second effect is that the shape of the wave front reflected from the grating repeats the shape of the grating but the proportionality coefficient is frequency dependent (see (5)). This effect is present in all four gratings and, to the best of our knowledge, has not been studied in the literature before.

3. The focal intensity is most strongly affected by the linear angular chirp (linear dependence of wave vector direction on frequency); this chirp occurs in both planes, even in a plane symmetric Treacy compressor. The decrease in the focal intensity associated with this chirp can be completely eliminated by rotating the G4 grating by an optimal angle. It has been shown that this simple method of suppressing space-time coupling can significantly reduce the requirements for the quality of grating surfaces. In practice, however, this will require measuring the pulse width at the focal point, since the minimum pulse width in the near field cannot serve as a feedback for the specified G4 grating alignment, or for the dispersion management, for example, for generating an AOPDF signal.

4. It has been shown that the decrease in focal intensity depends on the product of the grating surface rms $\sigma$ and the spectrum bandwidth $\Delta\omega$. For a given value of $\sigma$, with increasing $\Delta\omega$, the focal intensity changes in two ways: it increases in proportion to $1/\Delta\omega$ due to the shortening of the Fourier-limited pulse, but decreases due to an increase in space-time coupling (Fig. 4). As a result, if the Strehl ratio $St$<<1, a decrease in $\Delta\omega$ does not reduce the focal intensity, and may even lead to its slight increase. This circumstance can be used when designing not only a compressor, but also a laser as a whole.

The influence of space-time coupling in the stretcher on focal intensity is beyond the scope of this work. The beam size $w$ in the stretcher is significantly smaller than in the compressor, which influences the considered effects in different ways. On the one hand, they decrease, since the rms of the surface shape $\sigma$ is significantly less, because the rms is calculated over the beam aperture. On the other hand, they increase because the parameter $L_2/w$ is much larger. Thus, from (43), (42) it follows that the second effect described above in item 2 can be neglected for the stretcher, but the first effect will have the order of magnitude same as in the compressor.

We have considered space-time coupling caused by the phase part of the coefficient of reflection from the gratings. The proposed analytical approach can be generalized to the case of the amplitude part associated, for example, with beam clipping due to the fact that some rays at the edge of the spectrum go outside the aperture of the diffraction gratings. This effect has been studied in the literature only numerically [11]. In addition, the problem can be generalized to the case of non-ideal periodicity of the grooves, as well as their tilt.

The presented results have been obtained neglecting diffraction, which is justified for large-scale aberrations of diffraction gratings. Studying the influence of small-scale aberrations without allowance for diffraction is irrelevant and requires other approaches, which will be the subject of a separate publication.

# Appendix

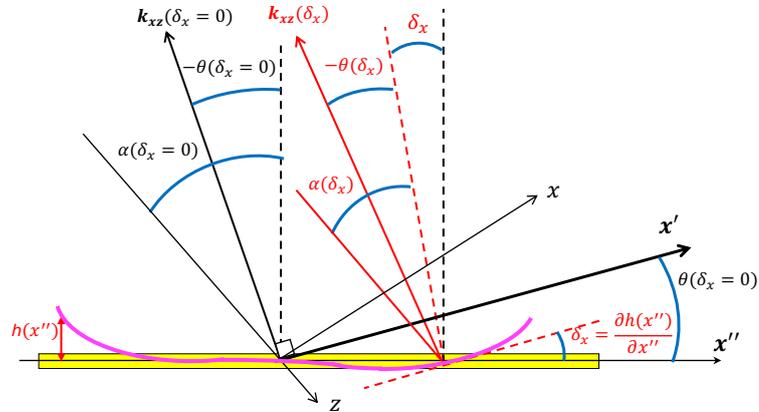

Consider a model of monochromatic beam reflection from the grating G1 or G3 with surface shape $h(x'', y'')$. We choose an arbitrary point on the grating surface with coordinates $(x'', y'')$. For the beam reflected from this point, we will find the wave vector $\mathbf{k}_{xz}(\delta_x)$ projection on the $x'$ axis normal to the vector $\mathbf{k}_{xz}(\delta_x = 0)$, i.e. normal to the wave vector of the reflected beam at $\delta_x = 0$. This projection is $\frac{\partial \varphi(x')}{\partial x'}$. Knowing it, we can find the phase for the incident beam $\varphi(x, y)$. It should be taken into account that, when reflected from the grating, the beam changes its size as $x = \frac{\cos\alpha}{\cos\theta(\omega)} x'$. Using the formula for the grating (6) we obtain

$$\varphi(x,y) = -\frac{\omega}{c}\cos\gamma\,(cos\theta(\omega) + cos\alpha)h\left(\frac{x}{cos\alpha}; \frac{y}{cos\gamma}\right) \qquad (A1)$$

For the mirror, $\theta(\omega) = \alpha$ and we can obtain the known expression $\varphi(\omega, \mathbf{r}) = -\frac{\omega}{c}2\cos\gamma\,cos\alpha\,h\left(\frac{x}{cos\beta}; \frac{y}{cos\gamma}\right)$. Note that this expression can be obtained for the Littrow angle for which $\theta(\omega) = -\alpha$. For the gratings G2 and G4, the angles $\alpha$ and $\theta(\omega)$ in (A1) change places and we will need a phase for the reflected beam $\varphi(x', y)$, so

$$\varphi(x', y) = -\frac{\omega}{c} \cos\gamma \left(\cos\theta(\omega) + \cos\alpha\right) h\left(\frac{x'}{\cos\alpha}; \frac{y}{\cos\gamma}\right)$$

Thus, we obtain

$$\varphi_n(x, y) = -\frac{\omega}{c} \cos\gamma_{1,2} \left(\cos\theta_{1,2}(\omega) + \cos\alpha_{1,2}\right) h_n\left(\frac{x}{\cos\alpha_{1,2}}; \frac{y}{\cos\gamma_{1,2}}\right),$$

where $n$ is the number of the grating, and the subscripts 1,2 of $\alpha, \theta, \gamma$ correspond to the first ($n = 1,2$) and second ($n = 3,4$) grating pairs.